\documentclass[reprint,pra,aps,amsmath,amssymb,floatfix,superscriptaddress,longbibliography]{revtex4-1}
\usepackage{physics}
\usepackage{dcolumn}
\usepackage{bm}
\usepackage{graphicx}
\usepackage{amsthm}
\usepackage{color}
\usepackage{physics}
\usepackage{xcolor}
\usepackage{verbatim}
\usepackage{esint}
\usepackage[english]{babel}
\usepackage{amsfonts}
\usepackage{slashed}
\usepackage{latexsym}
\usepackage{bm}
\usepackage[colorlinks = true,
            linkcolor = red,
            urlcolor  = blue,
            citecolor = magenta,
            anchorcolor = red]{hyperref}
\usepackage{esint}
\usepackage{soul}
\usepackage{cancel}
\usepackage[normalem]{ulem}
\usepackage{empheq}
\usepackage{dsfont}
\usepackage{amsmath}
\usepackage{ulem}
\usepackage{epsfig}
\usepackage{enumerate}
\definecolor{sapphire}{rgb}{0.03, 0.03, 0.41}
\DeclareMathAlphabet\mathbfcal{OMS}{cmsy}{b}{n}

\begin{document}

\title{Measurement-based quantum thermal machines with feedback control}

\author{Bibek Bhandari}
\affiliation{Institute for Quantum Studies, Chapman University, Orange, CA 92866, USA}
\author{Robert Czupryniak}
\affiliation{Department of Physics and Astronomy, University of Rochester, Rochester, NY 14627, USA}
\affiliation{Center for Coherence and Quantum Optics, University of Rochester, Rochester, NY 14627, USA}
\affiliation{Institute for Quantum Studies, Chapman University, Orange, CA 92866, USA}
\author{Paolo Andrea Erdman}
\affiliation{Freie Universit\"at Berlin, Department of Mathematics and Computer Science, Arnimallee 6, 14195 Berlin, Germany}
\author{Andrew N. Jordan}
\affiliation{Institute for Quantum Studies, Chapman University, Orange, CA 92866, USA}
\affiliation{Department of Physics and Astronomy, University of Rochester, Rochester, NY 14627, USA}

          
\begin{abstract}
We investigate coupled-qubit-based thermal machines powered by quantum measurements and feedback. We consider two different versions of the machine: 1) a quantum Maxwell's demon where the coupled-qubit system is connected to a detachable single shared bath, and 2) a measurement-assisted refrigerator where the coupled-qubit system is in contact with a hot and cold bath. In the quantum Maxwell's demon case we discuss both discrete and continuous measurements. We find that the power output from a single qubit-based device can be improved by coupling it to the second qubit. We further find that the simultaneous measurement of both qubits can produce higher net heat extraction compared to two setups operated in parallel where only single-qubit measurements are performed. In the refrigerator case, we use continuous measurement and unitary operations to power the coupled-qubit-based refrigerator. We find that the cooling power of a refrigerator operated with swap operations can be enhanced by performing suitable measurements.
\end{abstract}

\maketitle


\section{Introduction}
The quest to invent thermal machine at the nanoscale has led to the new field of quantum thermodynamics \cite{pekola2015,vinjanampathy2016,benenti2017,binder2019,landi2021,pekola2021}. Thanks to recent advances in nanofabrication techniques, much attention has been focused on realizing nanoscale-based quantum devices \cite{koski2014,koski2014demon,martinez2016,rossnagel2016,ronzani2018,josefsson2018,prete2019,maillet2019,senior2020} for heat management. Consequently, understanding how to control heat transport and dissipation at the nanoscale is of utmost significance and could enhance the performance of quantum devices' power and efficiency. Within the field of quantum thermodynamics, quantum thermal machines, such as heat engines and refrigerators, have been theoretically and experimentally investigated in detail \cite{scovil1959,geusic1967,alicki1979,pendry1983,geva1992,allahverdyan2000,schwab2000,kieu2004,meschke2006,pekola2007,blickle2012,horodecki2013,koski2013,brantut2013,thierschmann2015,anders2016,campisi2016,partanen2016,kosloff2017,tan2017,marchegiani2018,paolo2018,bibekgreen,bibekminimal,alonso2022}. 
Quantum refrigerators are quantum devices where heat is extracted from a cold thermal bath. Usually they are powered by external work provided by a chemical potential imbalance \cite{benenti2017,hajiloo2020}, or by external driving \cite{janinedot,erdman2019,brandner2020,bibekgeo,abiuso2,erdman2022}. 

Quantum limited measurements are now being performed regularly within the field of quantum computation. 
In contrast to classical measurement, quantum measurements can be ``invasive'', i.e. they can change the system's state and, consequently, the energetics of the system\cite{wiseman,jacobs,biele2017,buffoni2019}. This leads to a change in the quantum device's functioning and performance depending on the measurement type and strength\cite{caves1987,jacobs,weber2014,biele2017,wiseman,justin,lewalle2017,buffoni2019,monroe2021}. 
Particularly in the case of quantum devices, it can be important to keep track of the quantum measurement outcomes and act on the system accordingly to achieve a given task.

Technological advancement has enabled the
experimental realization of quantum thermal machines powered by measurements and feedback, such as Maxwell's demons\cite{maruyama2009,koski2014demon,cottet2017} and Szilard's engines\cite{koski2014}.
These are devices where measurements and feedback allow respectively the extraction of heat or work from a single thermal bath - apparently violating the second law of thermodynamics.
These realizations have motivated further research in the field, leading to an entire family of quantum measurement and feedback-based thermal machines. Heat and work extraction has been studied in various quantum systems exploiting quantum measurements with different strengths (weak or projective) and nature (invasive or non-invasive)\cite{jacobs2012,biele2017,buffoni2019,naghiloo2020,jayaseelan2021,bresque2021,manikandan2021efficiently,kagan,bhandari,yamamoto2022}. Although both invasive and non-invasive quantum measurements can be used to obtain information about the quantum system and run a feedback loop to power quantum thermal machines, it has been observed that invasive measurements alone can be used as fuel to power a thermal machine\cite{biele2017,buffoni2019,bhandari}.  

A Maxwell's demon powered by projective quantum measurements has been studied in single qubit systems in Refs.~\cite{pekola2016,elouard2017,cottet2017,miller2020,johnson2022}, and in double quantum dot systems in Ref.~\cite{annby2020}.
Recently, it was observed that also weak quantum measurements can be employed to realize a single qubit based Maxwell's demon and a refrigerator powered by invasive measurements and feedback\cite{kagan}. Furthermore, quantum measurements have also been utilized to realize heat engines\cite{jacobs2009,yi2017,chand2017,chand2018,ding2018,solfanelli2019,das2019,debarba2019,seah2020,hasegawa2020, anka2021, annby2022}, qubit elevators\cite{cyril2018}, quantum batteries\cite{gherardini2020,yao2022}, among other devices.

In this paper, we study various configurations of coupled-qubit-based thermal devices, namely a \textit{quantum Maxwell's demon}, and a \textit{measurement-assisted refrigerator}, the latter being a system that extracts heat from a cold bath exploiting the combination of external work and invasive quantum measurements.
As opposed to previous literature, we consider coupled-qubit-based devices powered by weak quantum measurements, both discrete and continuous. We study the performance of the machine in various configurations using different feedback strategies based on local measurements. In the Maxwell's demon case, we compare the impact of performing simultaneous measurements of both qubits on a single setup, and performing only individual qubit measurements on two setups operated in parallel. Thanks to a beneficial collective effect, we find that the former can outperform the latter. In the continuous measurement case, we compute the work distribution related to the stochasticity of measurement outcome, allowing us to observe quantities, such as power fluctuations, that are beyond average thermodynamic quantities. At last, in the refrigerator case we show how the addition of invasive quantum measurements, in the absence of feedback, can enhance the performance of a refrigerator powered by external work.  The results obtained in this paper for the case of measurement-assisted refrigerator can be straightforwardly extended to the case of coupled quantum dots attached to fermionic baths. In addition, the formulation used in this paper can be used to study finite-time statistics of different thermodynamic variables in terms of measurement record which can be directly accessible in an experiment\cite{kagan}.

The paper is organized as follows. In the next section, we introduce the models studied in this paper and the corresponding formalism. In Sec.~\ref{sec:demon}, we study the coupled-qubit device operated as a Maxwell's demon. We study both discrete and continuous measurements, as well as the impact of measuring a single qubit or both. In Sec.~\ref{sec:refrigerator}, we study the device operated as a measurement-assisted refrigerator under continuous measurements. In Sec.~\ref{sec:conclusion}, we draw the conclusions.

\section{Model}
\label{sec:model}
\begin{figure}[tb!]
\centering
\includegraphics[width=0.75\columnwidth]{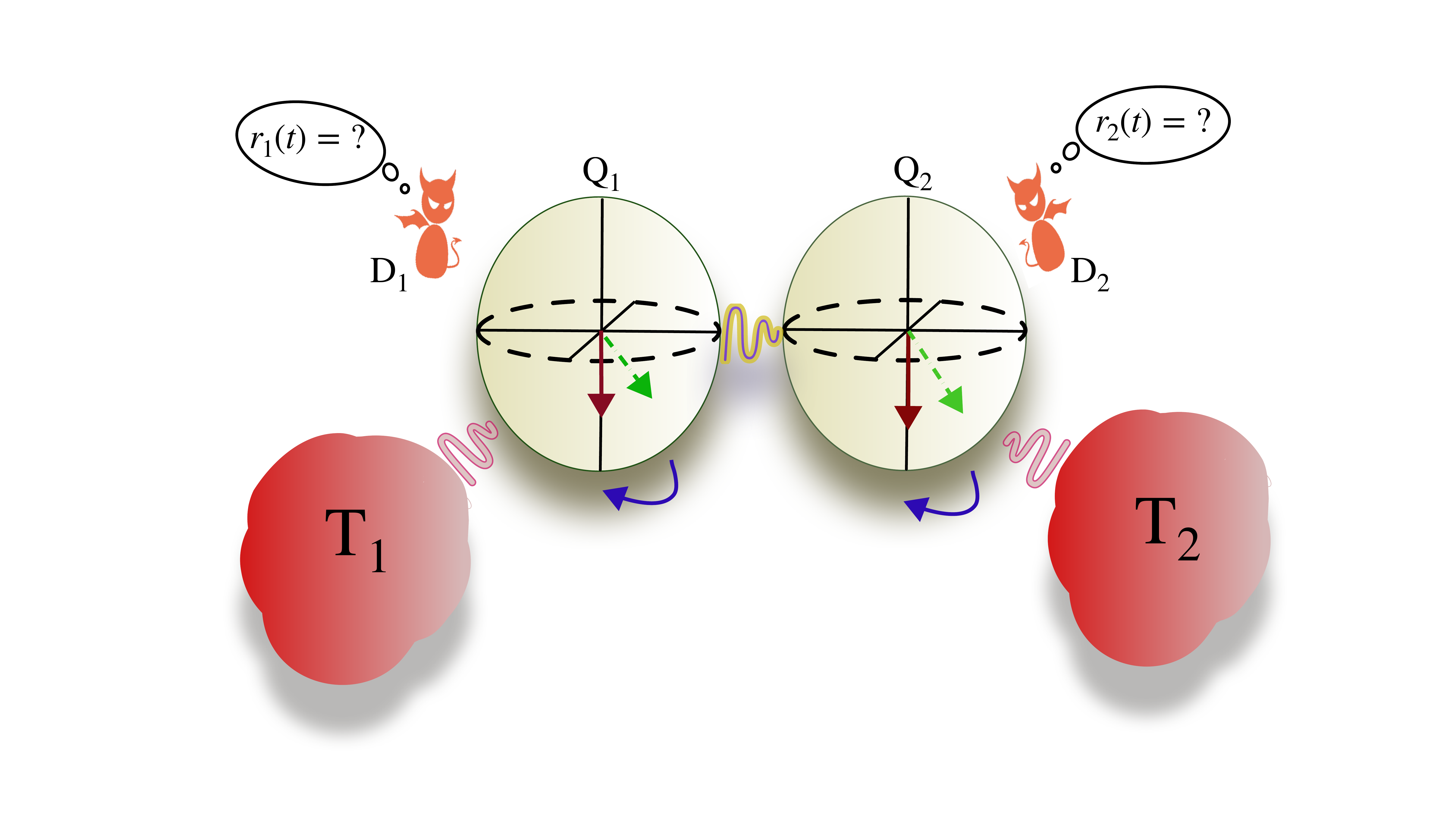}
 \caption{Coupled-qubit-based quantum feedback thermal machine. Qubit ${\rm Q}_i$ is attached to a thermal bath with temperature ${ T}_i$ and is being monitored by the measurement apparatus ${\rm D}_i$ for $i=1,2$. In the case of Maxwell's demon, $T_1=T_2=T$ whereas in the case of the measurement-assisted refrigerator, the two baths have different temperatures. Similarly, the two demons can undergo measurement with varying strengths. }
\label{fig:setup}
\end{figure}
We consider the setup of Fig.~\ref{fig:setup}: two coupled qubits, ${\rm Q}_1$ and ${\rm Q}_2$, are respectively coupled to two thermal baths at temperatures $T_1$ and $T_2$, and to two measurement apparatuses ${\rm D}_1$ and ${\rm D}_2$ that allow us to perform local quantum measurements on the respective qubits. The total Hamiltonian for the setup is given by $H = H_\text{Q} + H_\text{B} +H_\text{C}$, where the Hamiltonian of the coupled-qubit system is
\begin{multline}
H_\text{Q}=\frac{\epsilon_1}{2}\sigma_z^{(1)}+\frac{\epsilon_2}{2}\sigma_z^{(2)}+\Delta_x \sigma_x^{(1)}\sigma_x^{(2)}\\
+\Delta_y\sigma_y^{(1)}\sigma_y^{(2)} + \Delta_z \sigma_z^{(1)}\sigma_z^{(2)},
\label{eq:system_h}
\end{multline}
$\epsilon_i$  being the qubit gap for qubit ${\rm Q}_i$ and $\Delta_i$ the strength of the $\sigma_i$-$\sigma_i$ coupling between the two qubits. $H_\text{B}$ is the Hamiltonian describing the heat baths, which we consider as bosonic baths with continuous degrees of freedom 
\begin{equation}
H_\text{B} = \sum_{i=1,2}\sum_k\epsilon_{ik}b_{ik}^\dagger b_{ik},
\end{equation}
where $b_{ik}(b_{ik}^\dagger)$ are the bosonic annihilation (creation) operators with energy $\epsilon_{ik}$ and quantum number $k$ for bath $i$. We consider a linear ``tunnel-like'' coupling between the baths and the system given by
\begin{equation}
H_\text{C}=\sum_{i=1,2}\sum_k V_{ik}\left(\sigma_+^{(i)}b_{ik}+b_{ik}^\dagger \sigma_-^{(i)}\right),
\end{equation}
where $\sigma_{\pm}^{(i)}$ are ladder operators for qubit $Q_i$. 

We consider both discrete and continuous, as well as strong and weak quantum measurements. All such scenarios can be described by positive-operator-valued measures (POVMs), i.e. by a set of Krauss operators $M_k$, one for each measurement outcome, satisfying $\sum_k M_k^\dagger M_k = I$ for the discrete case and $\int dk M_k^\dagger M_k = I$ for the continuous case \cite{zhang2017,kagan}. The specific form of the Krauss operators for discrete and continuous measurements will be discussed in Sec.~\ref{sec:demon}. The probability (probability density in the continuous case) of measuring outcome $k$ is given by $\Tr[\rho M_k^\dagger M_k]$, where $\rho$ is the reduced density matrix of the coupled-qubit system. The post-measurement state $\rho_{\rm M_{k}}$, conditioned by observation $k$ and assumed to occur instantaneously, is given by
\begin{equation}
    \rho_{\rm M_{ k}} = \frac{M_k\rho M_k^\dagger}{\Tr[\rho M_k^\dagger M_k]}.
    \label{eq:state_meas}
\end{equation}

Throughout this paper, we consider two operational regimes: the \textit{quantum Maxwell's demon}, and the \textit{measurement-assisted refrigerator}. In the quantum Maxwell's demon case, we consider a single temperature of the environment, i.e. $T_1=T_2=T$. In this configuration, the aim is to extract heat from the single temperature bath exploiting invasive quantum measurements and feedback. In the measurement-assisted refrigerator, we consider an environment consisting of two different temperatures $T_1$ and $T_2$, and the aim is to maximize the heat extracted from the cold bath. Here the refrigerator is powered by a combination of work, delivered by an external control, and invasive quantum measurements in the absence of feedback.

\section{Quantum Maxwell's Demon}
\label{sec:demon}
In this section we describe our results operating the coupled-qubit-based thermal machine as a quantum Maxwell's demon. Here, we only consider the $\sigma_z$-$\sigma_z$ coupling between the two qubits, i.e. $\Delta_x=\Delta_y=0$.
In order to restrict the space of all possible quantum measurements and feedback strategies, we focus on local quantum measurements (i.e. using local probes ${\rm D}_1$ and ${\rm D}_2$ schematically shown in Fig.~\ref{fig:setup}), and local feedback strategies that are simple to implement experimentally. 

In particular, we consider unitary feedback consisting of local single-qubit unitary rotations around the y-axis, i.e. of the form $U_i(\theta_i)=e^{-i\theta_i\sigma^{(i)}_y}$, where $\theta_i$ is a suitable angle.
We consider both discrete and continuous quantum measurements of the spin state of each qubit in the ${\rm x}$-direction. Discrete weak $\sigma_x$ measurements performed on qubit ${\rm Q}_i$ using probe ${\rm D}_i$, for $i=1,2$, are described by the operators $\{M_{i +}, M_{i-}\}$, where
\begin{multline}
M_{i\pm} = \frac{1}{2}\left(\sqrt{\kappa_i}+\sqrt{1-\kappa_i}\right)I_2\otimes I_2
\\
\pm \frac{1}{2}\left(\sqrt{\kappa_i}-\sqrt{1-\kappa_i}\right) \cdot
\begin{cases}
\sigma_x^{(1)}\otimes I_2 \text{~ for $i=1$}, \\
I_2\otimes \sigma_x^{(2)} \text{~ for $i=2$} ,
\end{cases}
\end{multline}
$I_2$ is the 2x2 identity, and  $\kappa_i=1/2-\sqrt{2\gamma_i^{\prime}\delta t}$ is an indicator of the strength of the discrete measurement with characteristic measurement rate $\gamma^{\prime}_i$ and measurement time $\delta t$; these can be related to the resolution of the detector \cite{jacobs2003,wiseman}. The $k\to 0,1$ limits describe strong (projective) measurements, where the demon acquires maximum information about the system, whereas $k\to 1/2$ describes the opposite limit, where no information is acquired. Intermediate values of $k$ describe the transition from strong to weak measurements.

In the case of continuous measurement, we have a continuum of Krauss operators $\{{\cal M}_{i,r_i}\}_{r_i}$, one for each measurement apparatus $i$, where $r_i$ is the continuous measurement outcome.
They are given by
\begin{equation}
\begin{aligned}
&{\cal M}_{1,r_1}=\bigg(\frac{\delta t}{2\pi\tau}\bigg)^{\frac{1}{4}}\exp{-\frac{\delta t\left(r_1 I_2 -\hat{\sigma}_{x}^{(1)}\right)^2\otimes I_2}{4\tau}}, \\
&{\cal M}_{2,r_2}=\bigg(\frac{\delta t}{2\pi\tau}\bigg)^{\frac{1}{4}}\exp{-\frac{\delta t\,I_2\otimes\left(r_2 I_2 -  \hat{\sigma}_{x}^{(2)}\right)^{2}}{4\tau}},
\end{aligned}
\label{eq:m_continuous}
\end{equation}
where $\delta t$ is the time allocated to perform a single measurement and $\tau$ is the characteristic measurement time scale taken to separate the two measurement distribution by two standard deviations\cite{justin,kagan}. In other words, $\tau$ can be understood as the inverse of measurement strength and is the time required to achieve unit signal to noise ratio\cite{justin}. When $\delta t/\tau$ is large, the measurement is often referred to as strong measurement whereas the measurements for which $\delta t/\tau$ is small are called weak measurements. Following Eq.~(\ref{eq:m_continuous}), the measurement readout is randomly sampled from  two Gaussian distributions with variance $\sqrt{\tau/\delta t}$ and mean $+1$ (associated with the $\sigma_x=+1$ measurement outcome) and $-1$  (associated with the $\sigma_x=-1$ measurement outcome).

We operate the system as a Maxwell's demon considering the following thermodynamic cycle, consisting of three strokes: i) measurement, ii) feedback, and iii) thermalization.

i) Assuming the system to be initialized in a thermal state $\rho_{\rm T}=e^{-H_\text{Q}/(k_BT)}/Z$, where $Z=\Tr[e^{-H_\text{Q}/(k_BT)}]$, the initial energy of the coupled-qubits is given by
\begin{equation}
E_{\rm T} = {\rm Tr}\left[ \rho_{\rm T} H_Q\right].
\end{equation}
Let $\rho_i = \Tr_{\tilde{i}}[\rho]$ be the single-qubit density matrices given by tracing out the other qubit, where $\tilde{i}=2$ ($\tilde{i}=1$) for $i=1$($i=2$). 
Notice that, in the thermal state $\rho_{\rm T}$, the Bloch vectors of each single-qubit density matrix
only has a $z$ component, since $\Delta_x=\Delta_y=0$. A quantum measurement is now performed using either ${\rm D}_1$, or both ${\rm D}_1$ and ${\rm D}_2$. After performing a measurement, the state changes to $\rho_{\text{M}_\text{k}}$. Now the Bloch vector of the measured qubits acquires an $x$ component, and the norm of the vector may change.

ii) Feedback is performed by applying unitary rotations $U_i(\theta_i)$ around the y-axis to the qubits that have been measured. The angle $\theta_i$ is conditioned on the measurement outcome. Indeed, it is chosen such that the single-qubit states $\rho_i$, corresponding to the measured qubits, are rotated to the positive or negative z-axis of the Bloch sphere. The feedback that brings the state of $Q_i$ back to the positive (negative) z-axis will be denoted as ${\rm F}_i = 1$ (${\rm F}_i = -1$). The state after the measurement and feedback is given by $\rho_{\text{F}_\text{k}} = U_1(\theta_1) \rho_{\text{M}_\text{k}} U_1^\dagger(\theta_1)$ if only ${\rm D}_1$ is used, and by $\rho_{\text{F}_\text{k}} = U_2(\theta_2)U_1(\theta_1) \rho_{\text{M}_\text{k}} U_1^\dagger(\theta_1)U_2^\dagger(\theta_2)$ if both detectors are used. The energy of the system after measurement and feedback is given by $E_{\text{F}_{k}}=\Tr[\rho_{\text{F}_\text{k}}H_\text{Q}]$. Notice that, $E_{\text{F}_{+}}=E_{\text{F}_{-}}=E_{\text{F}}$.

iii) The cycle is closed allowing a full thermalization of the system with the thermal baths. During this stroke, the state of the system returns to $\rho_T$,  and an amount of heat $Q=E_T-E_\text{F}$ is extracted from the bath.

\subsection{Discrete one qubit measurement}
\label{subsec:one_qubit}
In this subsection we only perform measurements with ${\rm D}_1$. As a consequence, only the state $\rho_1$ of ${\rm Q}_1$ changes. Let us denote with $x_1$ and $z_1$ the x and z components of the Bloch vector corresponding to $\rho_1$ after the measurement. The angle of the unitary rotation $U_1(\theta_1)$ corresponding to feedback ${\rm F}_1=1$ is given by $\theta_1=-\frac{1}{2}\tan^{-1}\left(\frac{x_1}{z_1}\right)$, whereas for feedback ${\rm F}_1=-1$ it is given by $\theta_1=-\frac{1}{2}\tan^{-1}\left(\frac{x_1}{z_1}\right)+\pi/2$. The angle is chosen to rotate the qubit to the positive (${\rm F}_1=1$) or negative (${\rm F}_1=-1$) z-axis.
\begin{figure}[tb!]
\centering
\includegraphics[width=0.9\columnwidth]{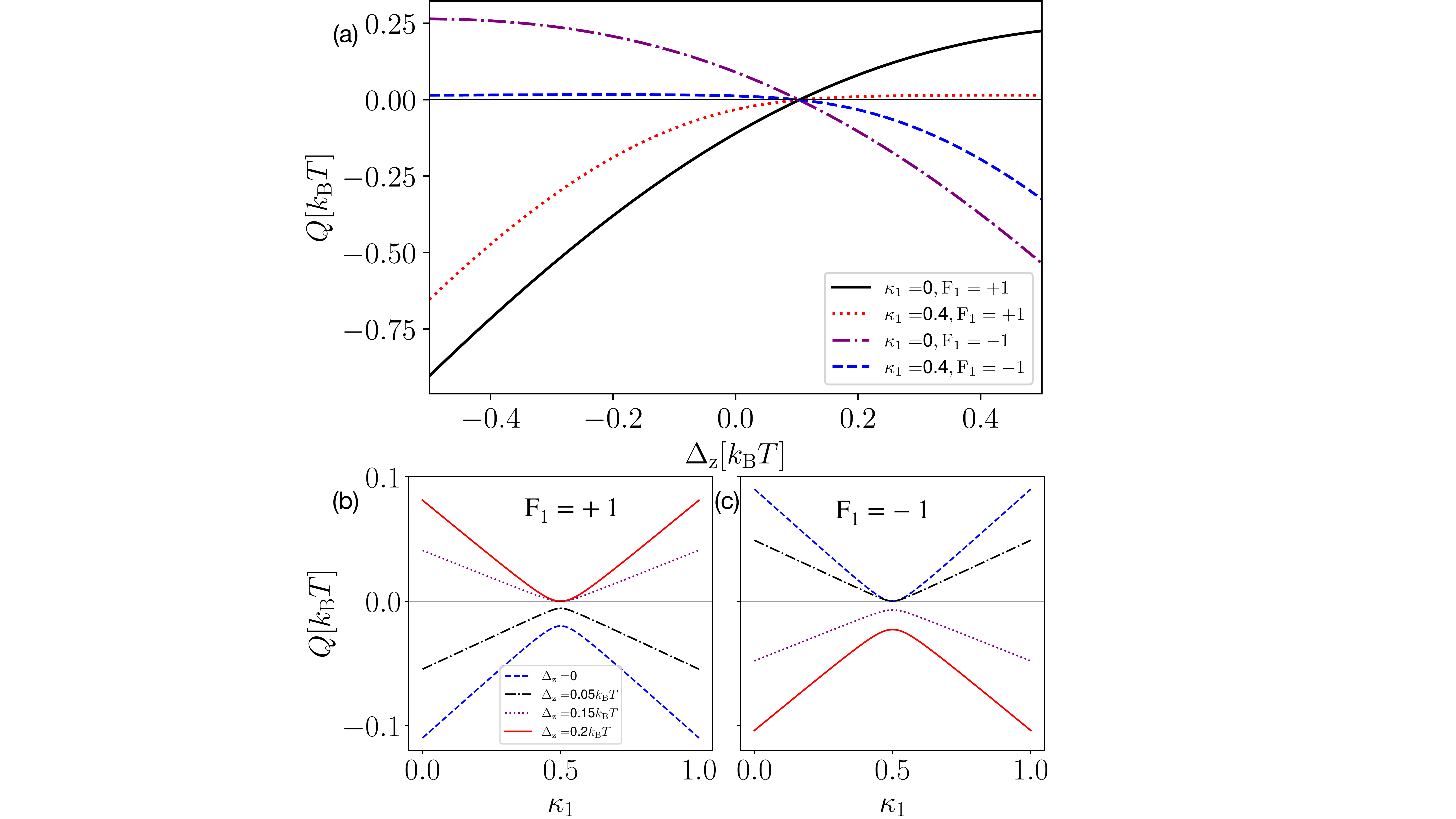}
	\caption{Heat extracted $(Q)$ as a function of qubit-qubit coupling strength (panel (a)) and measurement strength $\kappa_1$ (panel (b) for ${\rm F}_1=+1$ and panel (c) for ${\rm F}_1=-1$). In panel (a), the black and red curves give the heat extraction for two values of the measurement strength $\kappa_1$ and for feedback ${\rm F}_1=+1$ (rotation to positive z-axis). Similarly, the purple and blue curves give the heat extraction for feedback ${\rm F}_1=-1$ (rotation to negative z-axis). In panel (b) and (c), we plot the heat extraction as a function of $\kappa_1$ for ${\rm F}_1=+1$ and ${\rm F}_1=-1$ respectively taking different coupling strengths between the qubits. We take, $\epsilon_1=0.1 k_{\rm B}T$, $\epsilon_2=2k_{\rm B}T$. }
	\label{fig:diskappa}
\end{figure}

In Fig.~\ref{fig:diskappa} we investigate the heat $Q$, extracted from the heat bath, as a function of qubit-qubit coupling strength $\Delta_z$ for different values of $\kappa_1$ and feedback strategies (panel (a)), and as a function of the measurement strength $\kappa_1$ for different values of $\Delta_z$ (panel (b) corresponding to feedback ${\rm F}_1=1$ and panel (c) to ${\rm F}_1=-1$). In the case of decoupled qubits, i.e. $\Delta_z=0$, we know that heat extraction can be obtained only with ${\rm F}_1=-1$\cite{kagan}, since ${\rm F}_1=1$ would increase the energy of the qubit, resulting in heating the baths, rather than cooling them. However, for finite $\Delta_z$, we observe that positive heat extraction can be obtained even with ${\rm F}_1 = +1$ (see solid black and dotted red curves in Fig.~\ref{fig:diskappa}(a)). The heat extraction, in this case, is obtained when $\Delta_z > 0.1 k_{\rm B}T$. The behavior above can be explained by considering  that the energetics of the coupled systems is influenced by $\Delta_z$. Note that, for suitable choice of feedback, the heat extraction is an increasing function of $\Delta_z$ in the considered parameter regime. In Fig.~\ref{fig:diskappa}(b) and \ref{fig:diskappa}(c), we observe that for $\kappa_1=0.5$, i.e. when the demon acquires no information from the measurement, the qubit dissipates heat to the bath for all values of $\Delta_z$. Since no information is obtained, the demon has no resources to extract heat from a single thermal bath. Conversely, maximum heat is extracted from the bath when $\kappa_1\rightarrow 0,1$, which corresponds to maximum information extraction (the demon performs a projective measurement and feedback). Comparing Fig.~\ref{fig:diskappa}(b) and \ref{fig:diskappa}(c), we observe that changing the feedback strategy ${\rm F}_1$ from +1 (panel (b)) to -1 (panel (c)) or vice-versa changes the sign of heat extraction.
\subsection{Discrete two qubit combined measurement}
\begin{figure}[htb!]
\centering
\includegraphics[width=0.75\columnwidth]{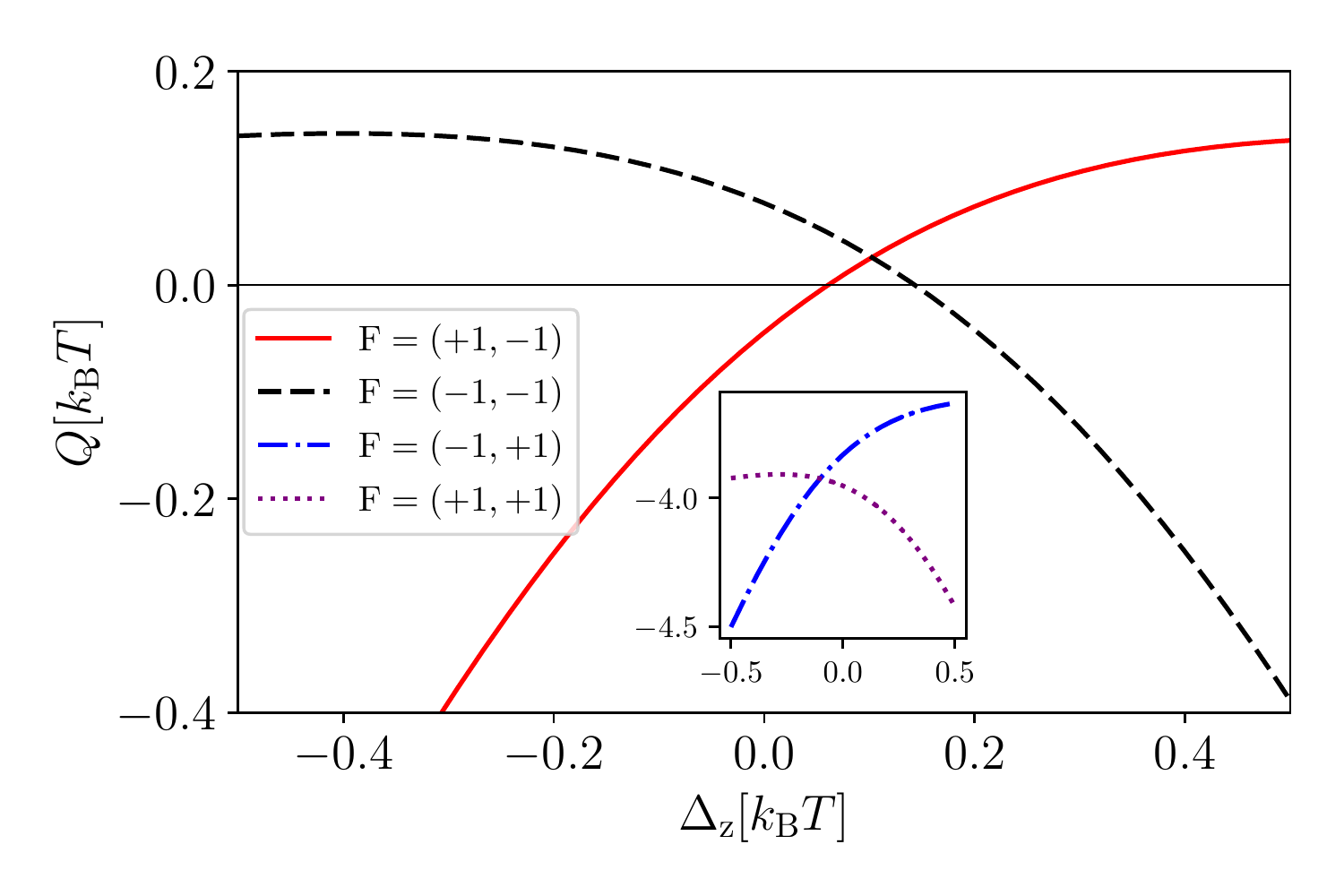}	
	\caption{Heat extracted $Q$ as a function of $\Delta_z$ for $\kappa_1=\kappa_2 = 0.2$. The feedback is represented as ${\rm F}=( {\rm F}_1, {\rm F}_2 )$, where ${\rm F}_i$ is the feedback applied to the qubit ${\rm Q}_i$. Finite heat extraction is obtained only when ${\rm F}_2 = -1$ (see the dashed black and solid red curves obtained with feedback ${\rm F}=(+1,-1)$ and ${\rm F}=(-1,-1)$ respectively). The dotted purple and dashed blue curves obtained with feedback $\text{F}=(+1,+1)$ and $\text{F}=(-1,+1)$ lead to the heating of the baths (see the inset). We take the same parameters as 
	Fig.~\ref{fig:diskappa}.}
	\label{fig:disdelta}
\end{figure}

\begin{figure}[htb!]
\centering
\includegraphics[width=0.75\columnwidth]{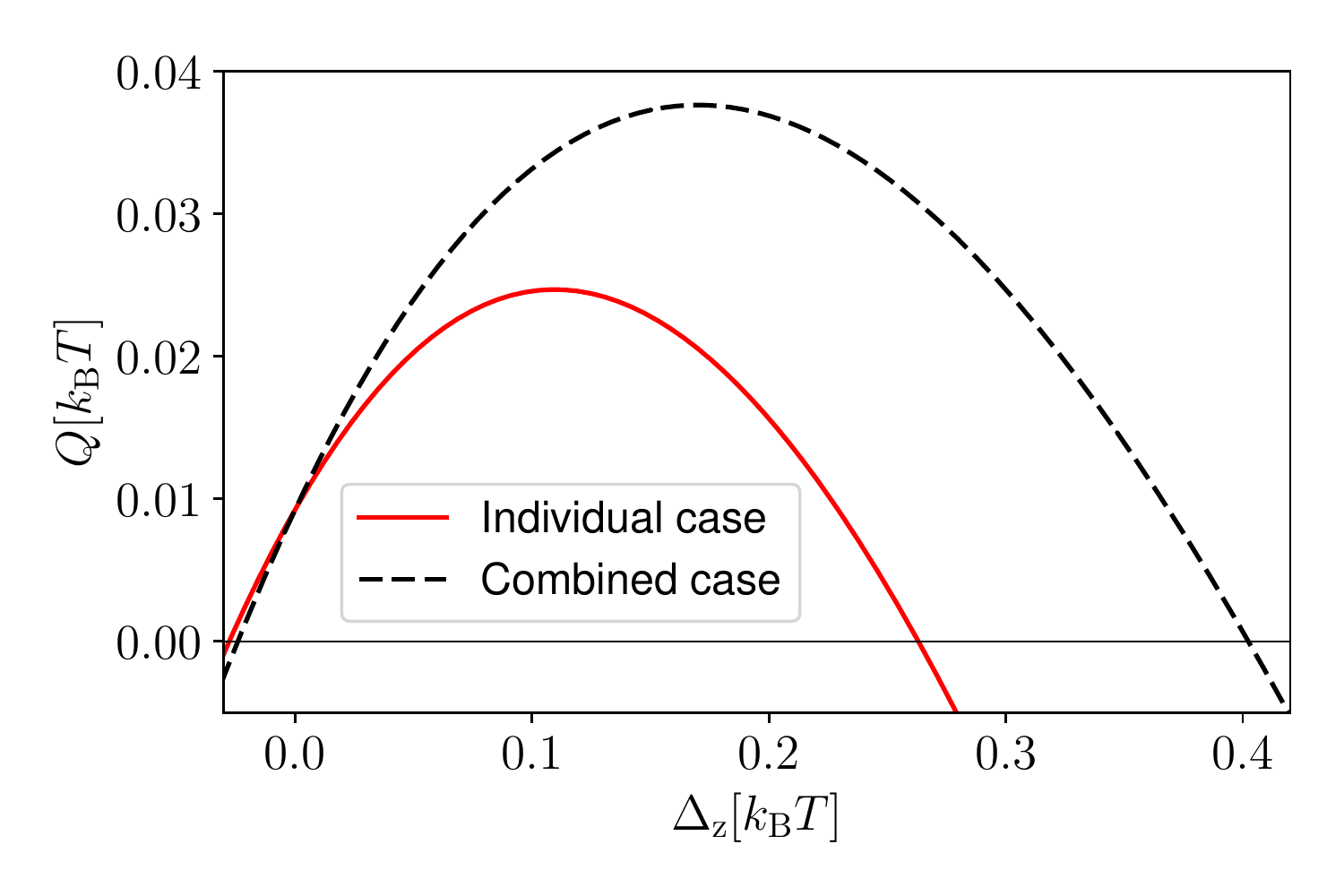}	
	\caption{Heat extracted ($Q$) as a function of $\Delta_z$ for individual measurement (solid red curve) and combined measurement of two qubits (dashed black curve). As feedback, we applied ${\rm F}=(+1,-1)$ in both cases. We take, $\epsilon_1 = 0.1k_{\rm B}T$, $\epsilon_2=0.5k_{\rm B}T$, $\kappa_1=\kappa_2 = 0.3$. }
	\label{fig:com_sin_sim}
\end{figure}
In this section we measure the state of the system using both ${\rm D}_1$ and ${\rm D}_2$ simultaneously. As a consequence, both $\rho_1$ and $\rho_2$ are affected by the measurement, so we will apply both $U_1(\theta_1)$ and $U_2(\theta_2)$ as feedback. Let us denote with $x_i$ and $z_i$ the x and z component of the Bloch vector corresponding to $\rho_i$ after measurement. The angle of the unitary operation that correspond to feedback ${\rm F}_i=1$ applied to ${\rm Q}_i$ is given by $\theta_i=-\frac{1}{2}\tan^{-1}\left(\frac{x_i}{z_i}\right)$, whereas for feedback ${\rm F}_i=-1$ applied to ${\rm Q}_i$ it is given by $\theta_i=-\frac{1}{2}\tan^{-1}\left(\frac{x_i}{z_i}\right)+\pi/2$. 

In Fig.~\ref{fig:disdelta} we study the heat extraction $(Q)$ out of the baths as a function of qubit-qubit coupling strength $(\Delta_z)$. Since $\epsilon_1\ll\epsilon_2,k_BT$ , we observe that heat extraction is possible only for the feedback ${\rm F}_2=-1$ on the qubit ${\rm Q}_2$ (see negative values of $Q$ in the inset). However, the choice of feedback on the qubit ${\rm Q}_1$ depends on the value of $\Delta_z$ (see red and black curves). As opposed to the single qubit case\cite{kagan} (see Fig.~\ref{fig:diskappa}a), here there are value of $\Delta_z$ where both feedback strategies ${\rm F}_1=+1$ and ${\rm F}_1=-1$ result in cooling. In the other limit when $\epsilon_2\ll\epsilon_1,k_{\rm B}T$ (not shown in the figure), heat extraction can be obtained only with feedback ${\rm F}_1=-1$ on ${\rm Q}_1$. For $\epsilon_1,\epsilon_2 \gg k_{\rm B}T$, we observe heat extraction only for the feedback ${\rm F}=({\rm F}_1,{\rm F}_2)=(-1,-1)$, since the system effectively behaves as two decoupled qubits.

We now study whether the combined use of both detectors ${\rm D}_1$ and ${\rm D}_2$ on a single coupled-qubit system (``combined case'') can lead to a better performance  with respect 
to having two coupled-qubit systems operated in parallel where only $D_1$ is applied to one system, and $D_2$ to the other one (``individual case'').
Notice that, in this comparison, the number of measurements is the same.
In Fig.~\ref{fig:com_sin_sim}, we plot the extracted heat as a function of $\Delta_z$, comparing these two scenarios.  The solid red curve corresponds to the individual case, whereas the dashed black curve corresponds to the combined case.
Notably, for the set of parameters considered, we observe that the combined case can outperform the individual case. Interestingly, we notice that the advantage of the combined case, i.e. the difference between the two curves, is enabled by the interaction between the qubits and, for $\Delta_z> 0$, it increases monotonically with increasing interaction strength. For large values
of $\Delta_z$ , the state of the coupled qubit system after feedback have larger
energy compared to its initial thermal energy leading to heating effect
instead of cooling.
\subsection{Continuous one and two qubit measurement}
Continuous feedback, which is widely used in optimal control in classical systems, depends on the continuous input of the measurement record. Continuous quantum measurement based feedback is a natural extension of the classical optimal control theory. In the quantum feedback theory based on continuous measurement, one studies the evolution of the density matrix under the influence of measurement and other external probes and suitably tunes the feedback control based on the continuous stream of measurement record. The evolution of the density matrix based on the stream of measurement record is referred to as a quantum trajectory\cite{carmichael2009,murch2013}. Experiments utilizing continuous measurement based feedback have been realized in several platforms, including quantum optics\cite{vijay2012} and quantum error correction\cite{livingston2022}.
\begin{figure}[htb!]
\centering
\includegraphics[width=0.9\columnwidth]{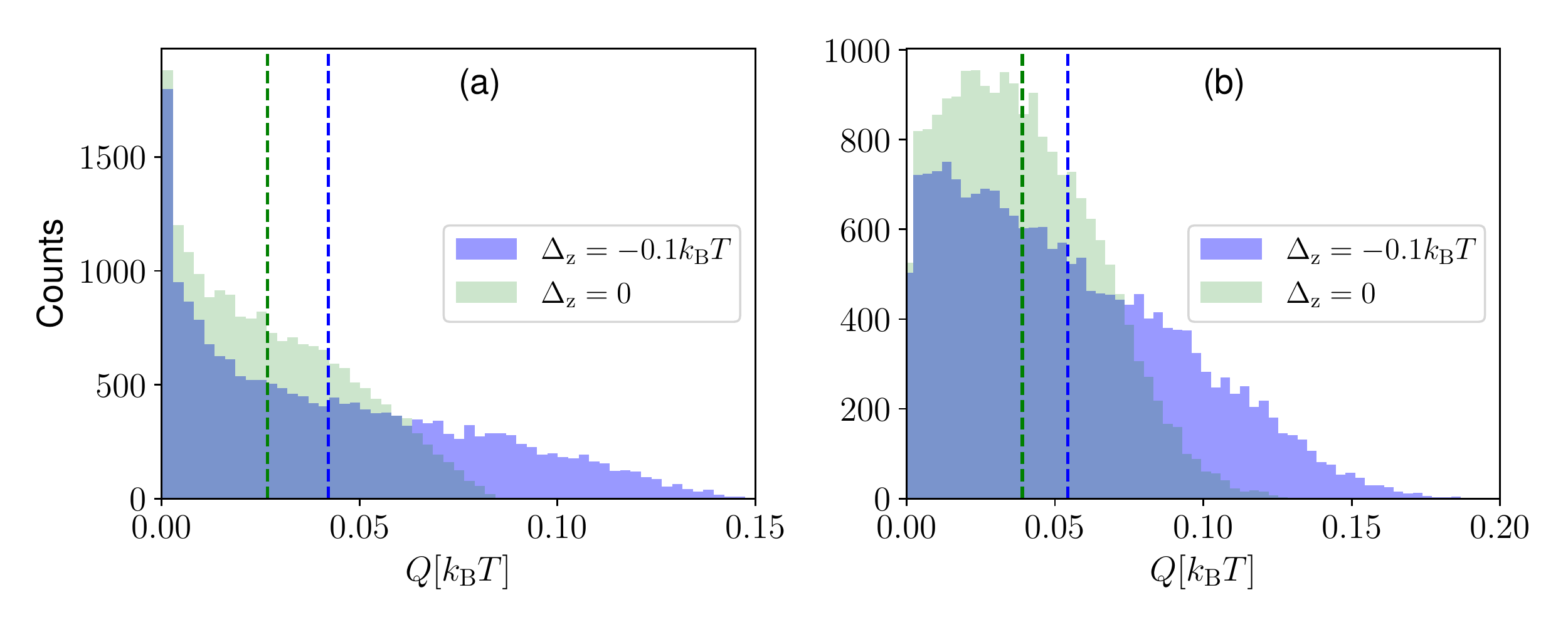}
	\caption{Count distribution of the heat extraction for one qubit continuous measurement (panel (a)) and the two qubit combined continuous measurement (panel (b)) for $
\delta t/\tau = 0.01$. The dashed lines indicate the averages of the distributions. The simulation is done for $n=20$ sequential measurements with feedback application only at the end. The distributions are for $N=20,000$ simulations.  As feedback, we applied ${\rm F}_1=-1$ in the left panel, and ${\rm F}=(-1,-1)$ in the right panel. We take the same parameters as 
	Fig.~\ref{fig:diskappa} for $\epsilon_1,\epsilon_2,k_{\rm B}T$. }
	\label{fig:conhis}
\end{figure}
\begin{figure}[htb!]
\centering
\includegraphics[width=0.9\columnwidth]{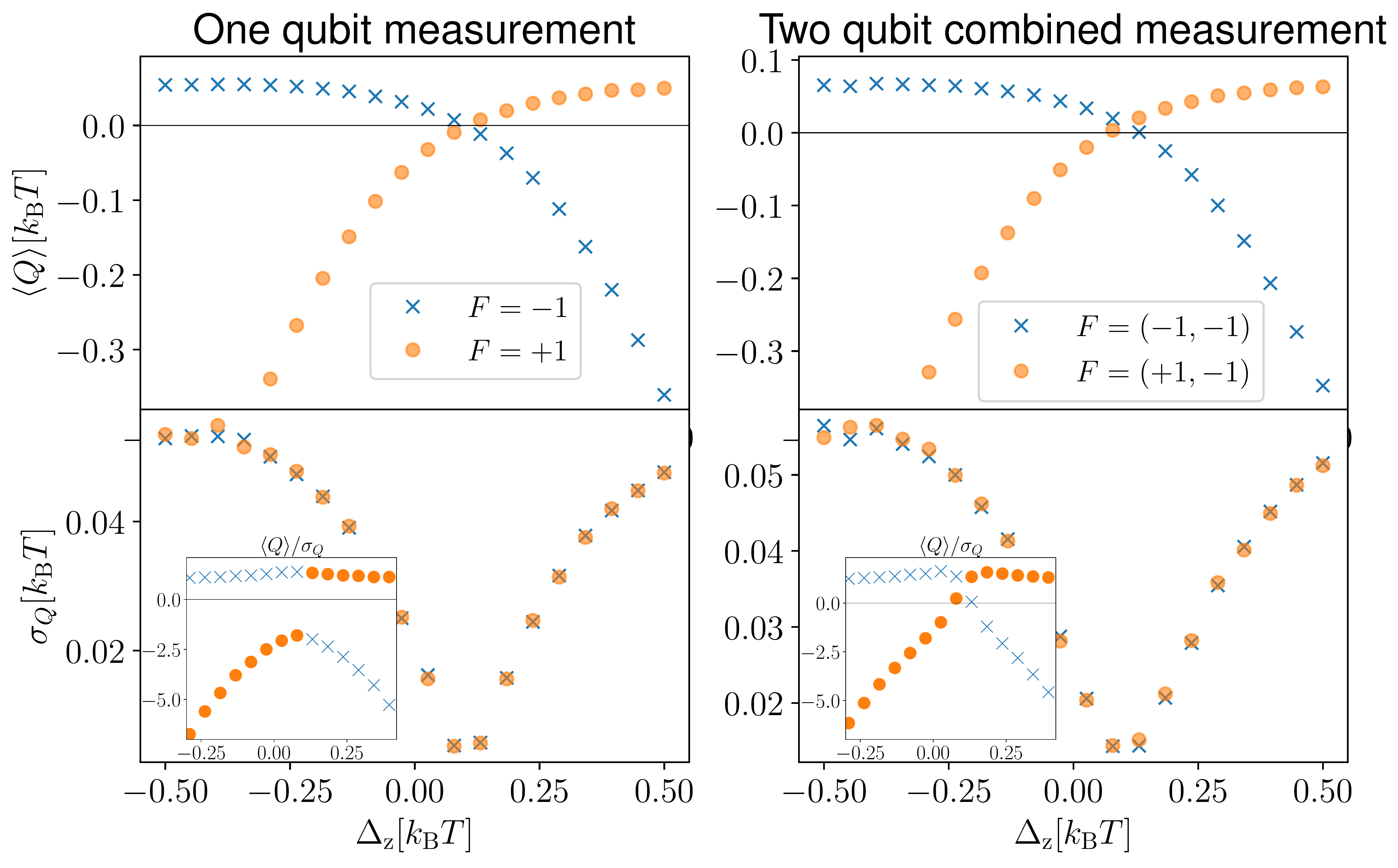}
	\caption{Average and standard deviation of the heat extraction for one qubit continuous measurements (left panels) and two qubit combined continuous measurements (right panel) for $\delta t/\tau = 0.01$. The simulation is done for $n=20$ sequential continuous measurements with feedback application only at the end. The distributions are for $N=20,000$ simulations. As feedback, we applied ${\rm F}_1=-1$ in the left panel, and ${\rm F}=(-1,-1)$ in the right panel.  In the inset, we show the variation of the signal (average heat extracted) to noise (standard deviation of the extracted heat) ratio as a function of $\Delta_z$. We take the same parameters as 
	Fig.~\ref{fig:diskappa} for $\epsilon_1,\epsilon_2,k_{\rm B}T$.}
	\label{fig:conmean}
\end{figure}
In this subsection, we study the distribution of the extracted heat performing a cooling cycle as in the previous subsections, but replacing the discrete measurement with a continuous measurement. Using Eq.~(\ref{eq:state_meas}), the state of the coupled-qubit system after measurement can be written as
\begin{equation}
\rho_{{\cal M},{r_1}}=\frac{{\cal M}_{1,r_1} \rho {\cal M}_{1,r_1}^\dagger}{\Tr[\rho {\cal M}_{1,r_1}^\dagger {\cal M}_{1,r_1}]},
\label{eq:den_meas1}
\end{equation}
when measurement is performed only with ${\rm D}_1$
and 
\begin{equation}
\rho_{{\cal M},{r_1r_2}}=\frac{{\cal M}_{2,r_2}{\cal M}_{1,r_1}\rho {\cal M}_{1,r_1}^\dagger {\cal M}_{2,r_2}^\dagger}{\Tr[\rho {\cal M}_{2,r_2}^\dagger {\cal M}_{1,r_1}^\dagger {\cal M}_{1,r_1}{\cal M}_{2,r_2}]},
\end{equation}
for combined measurement.

We describe a continuous measurement as a sequence of $n$ measurements of duration $\delta t$, each one described by the Krauss operators in Eq.~(\ref{eq:m_continuous}). Each sequence of measurement produces a trajectory for the state of the coupled-qubit system. In order to calculate the average and variance of the heat extraction, we shall consider $N$ different trajectories. Along each trajectory, we compute the exchanged heat for that particular sequence of measurement outcomes taking into account only the stochasticity induced by the quantum measurements, and not by the stochastic nature of heat exchange with the baths\cite{hekking2013,naghiloo2020}.
\begin{figure}[htb!]
\centering
\includegraphics[width=0.75\columnwidth]{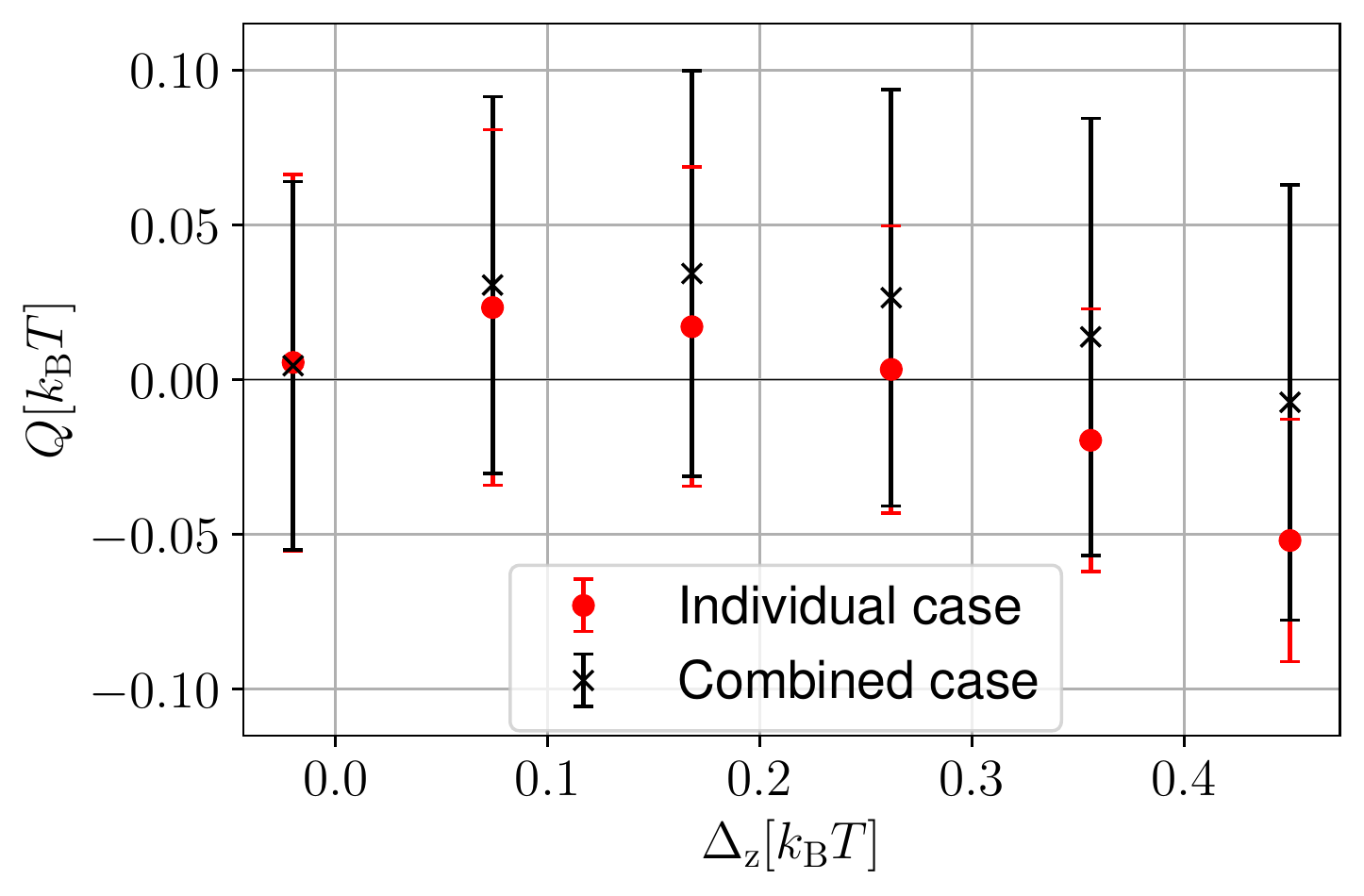}
	\caption{Heat extracted $Q$ as a function of $\Delta_z$ in the individual (solid red curve) and the combined (dashed black curve) cases. The circles and crosses represent the average heat extracted $\langle Q\rangle$ whereas the error bars give the respective fluctuations $\sigma_Q$. We consider the number of measurements $n=20$, and the number of trajectories $N=5000$. As feedback, we applied ${\rm F}=(+1,-1)$ in both cases. We take $\epsilon_1 = 0.1k_{\rm B}T$, $\epsilon_2=0.5k_{\rm B}T$, $\delta t/\tau =0.01$. }
	\label{fig:con_int}
\end{figure}

In Fig.~\ref{fig:conhis}, we compare the extracted heat distribution in the one qubit measurement case (panel (a)) and in the combined measurement case (panel (b)) for $N=20,000$ simulations of the
heat extraction processes and for ${\rm F}_1=-1$ in the left panel, and ${\rm F}=(-1,-1)$ in the right panel. Each simulation is obtained by
performing feedback after $n = 20$ sequential measurements each of duration $\delta t$. Interestingly, in the case of one qubit continuous measurement, we find that the engine is more likely to extract zero heat, and the probability of extracting heat $Q>0$ decreases monotonically with $Q$. However, in the combined measurement case, a finite amount of heat (whose magnitude depends on the value of $\Delta_z$) is extracted more often than zero heat. 
The distribution in green is for the case when the two
qubits are decoupled, whereas the blue distribution gives
the finite coupling case $(\Delta_z=-0.1 k_{\rm B}T)$. We observe
that the coupled system produces a larger average heat extraction (blue dashed line) compared to the decoupled
system (green dashed line). However, the greater average
heat extraction is accompanied by larger fluctuations, as observed from the broader width of the probability distribution for $\Delta_z = -0.1 k_{\rm B}T$. This can also be observed in Fig.~\ref{fig:conmean} where we plot the average heat extraction $\ev*{Q}$ (upper panels) and the
fluctuation quantified by the standard deviation $\sigma_Q$ (lower panels), as a function of $\Delta_z$. The panels on the left-hand side are for one qubit measurement case, whereas the
right-hand side corresponds to the two qubits combined
measurement case. We observe that the standard deviation reaches a minimum when the average extracted heat goes to
zero. In addition, fluctuations are present both when the system is cooling and heating the environment. The maximum
fluctuation is observed when the average heat extraction
takes the maximum value. From the inset, we observe that the ratio between average heat extracted and its standard deviation shows a maximum as a function of $\Delta_z$ in the $\langle Q\rangle >0$ regime. Comparing the one qubit measurement and two qubit combined measurement cases, we observe that although combined measurement gives better average heat extraction, it
is also associated with larger fluctuations. 

As we did for the discrete measurement case (see Fig.~\ref{fig:com_sin_sim}), we now we assess the impact of combined quantum measurements. In Fig.~\ref{fig:con_int} we compare the ``individual'' and ``combined'' cases. The average heat extraction in the individual case is denoted with red circles, whereas the combined case is given by black crosses. The errors bars denote the standard deviation in the respective cases. As we observed in the discrete measurement case, there are system parameters (as the ones chosen in Fig.~\ref{fig:con_int}) where the combined case outperforms the individual case. However, the larger heat extraction is also accompanied by larger fluctuations (compare the range of red and black error bars). This highlights once again the benefits of collective measurements for the average power of quantum thermal machines at the expense of larger fluctuations. This trade-off between power and power fluctuations is reminiscent of the thermodynamic uncertainty relations that have been derived, in the absence\cite{barato2015,gingrich2016,pietzonka2018,guarnieri2019,timpanaro2019,falasco2020,friedman2020,miller2021} and presence\cite{potts2019} of measurements, for quantum thermal machines.

\section{Measurement-assisted refrigerator}
\label{sec:refrigerator}
In this section we operate the continuously monitored coupled-qubit system as a measurement-assisted refrigerator. The two baths are considered non-detachable and will be kept at different temperatures to realize a refrigerator. More specifically, the refrigerator is powered by ``swap operations''\cite{campisi}, which can be interpreted as work provided by a time-dependent driving that implements the unitary swap operation, and by invasive quantum measurements in the absence of feedback. The state $\rho$ of the coupled qubits weakly coupled to the heat baths, under the influence of continuous measurements, is described by
\begin{equation}
\frac{d\rho}{dt}=-i\left[H_{\rm Q},\rho\right]
+{\cal L}_{\rm B} \rho+{\cal L}_M\rho,
\label{eq:meas_main}
\end{equation}
where $[\dots,\dots]$ represents the commutator. The first term in the right hand side represents the unitary evolution of the system, whereas ${\cal L}_\text{B}$ is a linear superoperator describing the dissipative dynamics induced by the coupling to the baths. To ensure thermodynamic consistency of our results, the dissipative term ${\cal L}_\text{B}\rho$ is derived using the global master equation\cite{breuer} that satisfies local detailed balance (see App.~\ref{app:mas_equ} for details).
This guarantees that, for $T_1=T_2=T$ and in the absence of measurements and feedback, the state will evolve into thermal Gibbs state $\rho_T$.
The third term on the right hand side of Eq.~(\ref{eq:meas_main}) is the quantum measurement contribution.  It can be expressed as\cite{jacobs,wiseman,zhang2017}
\begin{equation}
{\cal L}_M\rho=\Gamma_M{\cal D}[X]\rho 
+\sqrt{\Gamma_M}{\cal H}[X]\rho
\frac{dW}{dt},
\end{equation}
 where ${\cal D}[X]\rho =\left[X\rho X -\frac{1}{2}\left[XX\rho +\rho XX\right]\right]$ gives the dissipative contribution of quantum measurement and ${\cal H}[X]\rho=\left[X\rho +\rho X-2\langle X\rangle\rho\right]$ is the stochastic contribution. $\Gamma_M$ determines the strength of the measurement and $X$ is the system observable being measured. Only the first term survives upon averaging over ensemble of measurement records. $dW$ is a stochastic quantity which results from the random nature of measurements. The distribution for $dW$ is Gaussian with zero mean and variance $dt$. 

Let us denote the product of the eigenstates of $\sigma_z^{(i)}$ as $\{|0\rangle, |1\rangle, |2\rangle, |d\rangle\}$ where $|0\rangle$ represents the state where both qubits are in the ground state, $|1\rangle$ when only qubit 1 is excited, $|2\rangle$ when only qubit 2 is excited, and $|d\rangle$ when both qubits are excited. Motivated by the Hamiltonian of tunnel coupled single level quantum dot systems, where the doubly excited state may be energetically prohibited due to strong Coulomb interactions between the two quantum dots, we choose
 $E_1=(\epsilon_1-\tilde{\Delta})/2$, $E_2=(\epsilon_2-\tilde{\Delta})/2$, $\Delta_z = \tilde{\Delta}/2$, $\Delta_x=\Delta_y=\Delta/2$, and we consider the limit of large interaction $\tilde{\Delta}/(k_BT)$. We can thus neglect the $|d\rangle$ state, and Eq.~(\ref{eq:system_h}) for the coupled-qubit Hamiltonian reduces to
\begin{equation}
H_\text{Q}=E_1 |1\rangle\langle 1|+E_2|2\rangle\langle 2| 
+ \Delta(|1\rangle \langle 2| +|2\rangle\langle 1|).
\end{equation}
The diagonalization of $H_\text{Q}$ leads to the basis $\left\{|0\rangle,|+\rangle,|-\rangle\right\}$, where the energy of the state $|0\rangle$ is zero, and the energy of the states $|\pm\rangle$ is
\begin{equation}
E_{\pm} =\frac{E_1+E_2}{2}\pm \frac{1}{2}\sqrt{(E_1-E_2)^2+4\Delta^2}.
\end{equation}
The master equation in Eq.~(\ref{eq:meas_main}) prescribes the evolution for the density matrix which we express in the  $\left\{|0\rangle,|+\rangle,|-\rangle\right\}$ basis in terms of transition rates and measurement parameters (see Ref.~\cite{bhandari} for details). Here, we measure the state of $Q_2$ using $D_2$ by measuring the operator $\Pi_X=|2\rangle \langle 2|$.
\begin{figure}[htb!]
\centering
\includegraphics[width=0.75\columnwidth]{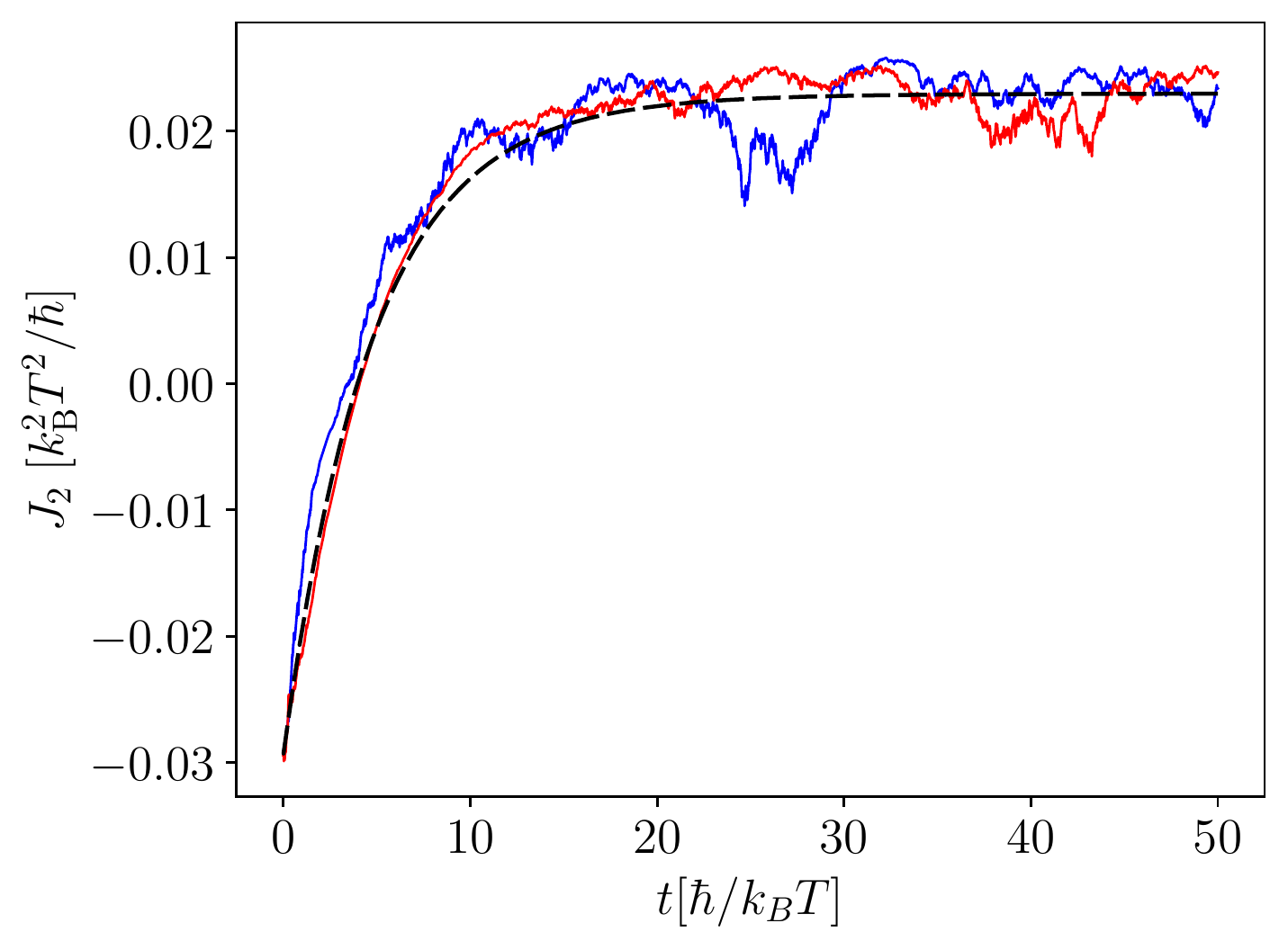}	
	\caption{Refrigeration obtained as a result of measurement and swap operation in the coupled-qubit system attached to two baths with different temeperatures. The black dashed line gives the average heat current whereas the red and blue curves are the heat current obtained for single trajectory of measurement. We take $E_1 = 5 k_{\rm B}T$, $E_2=2k_{\rm B}T$, $\Gamma_1=\Gamma_2=0.05$, $\Delta=0.2 k_{\rm B}T$, $\Gamma_M=0.02 k_{\rm B}T$, $T_1=1.1 T$, $T_2=T$, and $\delta t=0.01\hbar/k_{\rm B}T $. $\Gamma_i$ parameterizes the coupling strength between the coupled-qubit system and the bath $i=1,2$ (see App.~\ref{app:mas_equ}).}
	\label{fig:cool_swap}
\end{figure}
In addition, after every measurement step of duration $\delta t$ we apply a unitary rotation given by 
\begin{equation}
U_{\rm rot}=\begin{bmatrix}
1 & 0 & 0  \\
0 & \cos\Theta &\sin\Theta &\\
0 & \sin\Theta & -\cos\Theta
\end{bmatrix}.
\end{equation}
For $\Theta=\pi/2$, the unitary rotation $U_{\rm rot}$ becomes an effective swap gate $U_{\rm SWAP}$ between the $|+\rangle$ and $|-\rangle$ states.

Although we keep both diagonal and off-diagonal terms in our density matrix, we observe that in the weak coupling and weak measurement limit the contribution from the off-diagonal terms is very small compared to the contribution from the diagonal terms for $\Delta\ll E_1,E_2$.

In Fig.~\ref{fig:cool_swap}, we study the heat current $J_2$ flowing out of the bath at temperature $T_2\leq T_1$ when the coupled-qubit system is subject to continuous measurement and swap operation after each measurement. The average heat flow out of the colder bath is given by the dashed black curve. The blue and red curves are obtained when only individual trajectories are considered and takes into account the stochastic nature of measurement. We observe that when $E_1>E_2$, the swap operation leads to considerable cooling effect. However, when $E_2>E_1$, the swap operation leads to the heating effect.
\begin{figure}[htb!]
\centering
\includegraphics[width=0.9\columnwidth]{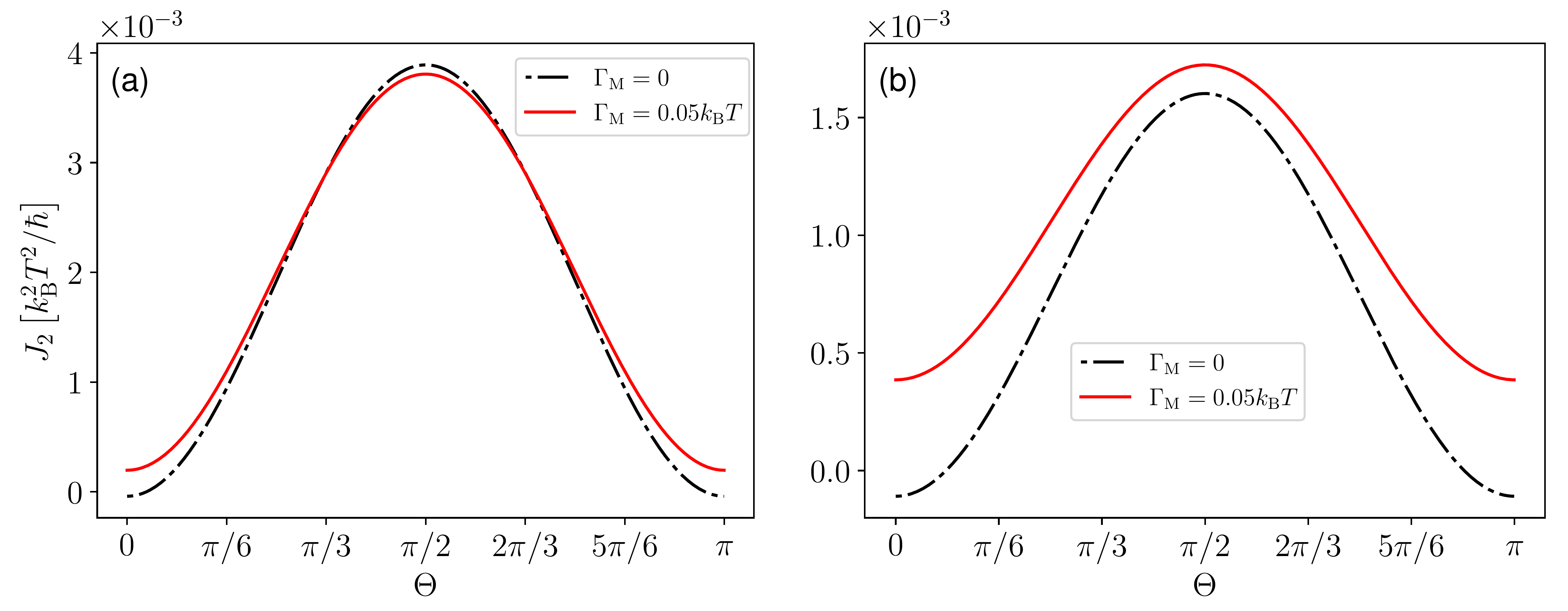}	
	\caption{Heat current flowing out of the cold bath as a function of the rotation angle $\Theta$ for $\Delta=0.5 k_{\rm B}T$ (panel (a)) and $\Delta = k_{\rm B}T$ (panel (b)). We take, $T_1=1.1 T$, $T_2=T$, $E_1 = 5 k_{\rm B}T$, $E_2 = 2 k_{\rm B}T$, $\Gamma_1=\Gamma_2=0.01$.}
	\label{fig:theta1}
\end{figure}

In Fig.~\ref{fig:theta1}(a), we study $J_2$ as a function of rotation angle $\Theta$ for $E_1>E_2$. We observe that cooling is obtained even when there is no input work ($\Theta=0$)\cite{bhandari} and maximum cooling is obtained for swap operation ($\Theta = \pi/2$). The former is possible thanks to the invasive nature of quantum measurements which changes the energetics of the quantum system leading to cooling effect upon an appropriate choice of the measurement\cite{biele2017,buffoni2019}.
The black dashed curve is obtained when there is input work without measurement and the red curve is obtained when one considers both measurement and input work. We observe that the impact of continuous measurements on the heat current can be positive or negative depending on the rotation angle $\Theta$.
However, changing the value of parameter $\Delta$, we observe in Fig.~\ref{fig:theta1}(b) that a parameter regime exists where the combined effect of the invasive measurement and external work gives a better cooling effect for all values of $\Theta$.

\section{Conclusions}
\label{sec:conclusion}
We studied coupled-qubit-based quantum thermal machines powered by quantum measurement and feedback. In the case of Maxwell's demon, we studied various ways of implementing quantum measurement. We investigated both discrete and continuous measurement as well as one qubit measurement and two qubit combined measurement. In the case of one qubit measurement, and for a suitable choice of feedback, we observed that the heat extraction from the thermal bath increases monotonously as a function of $\sigma_z$-$\sigma_z$ coupling strength ($\Delta_z$) between the two qubits for a range of values of $\Delta_z$. We then compare the heat extracted from a single setup subject to combined measurements of both qubits, with the heat extracted from two setups operated in parallel where only individual qubits are measured. Thanks to a collective effect, we find that the former can outperform the latter. In the case of continuous measurement, we studied the distribution of heat extraction for both one and two qubit measurement. Similar to the case of discrete measurement, in certain parameter regime, we observed better average heat extraction with combined measurement of two qubits compared to individual measurement of each qubit in two parallel setups. However, better average heat extraction was always associated with higher fluctuations. 

In the second part of the paper, we studied the measurement-assisted refrigeration in the coupled-qubit sytem attached to two thermal baths at different temperatures. We showed that although measurement and swap operations alone can power a refrigeration, a combination of the two can yield higher refrigeration.

\section{Acknowledgements}

This work was supported by the U.S. Department of Energy (DOE), Office of Science, Basic Energy Sciences (BES), under Award No. DE-SC0017890. PAE gratefully acknowledges funding by the Berlin Mathematics Center MATH+ (AA1-6).
\appendix

\section{Quantum master equation}
\label{app:mas_equ}
The contribution of the baths to the master equation can be written in terms of transition rates given by
\begin{align}
&{\Gamma}_{i,m0}=\hbar^{-1}\gamma_i(\epsilon_{m0})\Big(1\pm n_i(\epsilon_{m0})\Big),\nonumber \\
&{\Gamma}_{i,0m}=\hbar^{-1}\gamma_i(\epsilon_{m0})n_i(\epsilon_{m0}),
\end{align}
where $m=+,-$ are the states of the coupled-qubit system, $\gamma_i(\omega)= \Gamma_i \omega e^{-\omega/\omega_C}$ and  $n_i(\omega)=(\exp{\omega/k_{\rm B}T_i}-1)^{-1}$ are the Ohmic spectral density and the Bose-Einstein distribution function for the bath with temperature $T_i$, $\omega_C$ is the cut-off frequency. $\Gamma_{i,m0}$ gives the transition rate into the bath $i$ whereas $\Gamma_{i,0m}$ gives the outgoing transition rate from bath $i$. The full master equation with the contribution from the baths as measurement probe including both diagonal and off-diagonal terms has been studied in the appendix of Ref.~\cite{bhandari}.

\bibliography{references}

\end{document}